\documentclass[aps,pre,reprint,superscriptaddress,amsmath,showpacs,amssymb]{revtex4-2}
\usepackage{hyperref}
\usepackage{graphicx}
\usepackage{dcolumn}
\usepackage{url}
\usepackage{bm}

\usepackage{subfigure}
\usepackage{siunitx}
\usepackage{todonotes}
\usepackage[mathscr]{eucal}

\presetkeys{todonotes}{inline}{}

\usepackage{booktabs}

\newcommand{\df}{\mathrm{d}}
\newcommand{\vv}[1]{\bm{#1}}
\newcommand{\te}[1]{\text{#1}}

\newcommand{\ie}{i.e.}

\newcommand{\Dr}{{D_\te r}}

\newcommand{\Pa}{{p_\te a}}
\newcommand{\Pp}{{p_\te p}}
\newcommand{\Ra}{{\rho_\te a}}
\newcommand{\Rp}{{\rho_\te p}}
\newcommand{\la}{{\lambda_{\te a}}}
\newcommand{\lp}{{\lambda_{\te p}}}

\newcommand{\Ca}{{C_\te a}}
\newcommand{\Cp}{{C_\te p}}

\newcommand{\rif}{{r_\te{if}}}
\newcommand{\hn}{{\vv{\hat n}}}
\newcommand{\hr}{{\vv{\hat r}}}
\newcommand{\mynabla}{{\boldsymbol{\nabla}}}

\DeclareSIUnit\length{\text{$\sqrt{D/2k}$}}
\DeclareSIUnit\lengthcont{\text{$\sqrt{D/\Dr}$}}

\begin{document}

\preprint{APS/123-QED}

\newcommand\uleipT{\affiliation{Institute for Theoretical Physics, Leipzig University, 04103 Leipzig, Germany}}
\newcommand\chau{\affiliation{ 
 Charles University,  
 Faculty of Mathematics and Physics, 
 Department of Macromolecular Physics, 
 V Hole{\v s}ovi{\v c}k{\' a}ch 2, 
 CZ-180~00~Praha, Czech Republic 
}}

\title{Density and Polarization of Active Brownian Particles in Curved Activity Landscapes}

\author{Sven Auschra}
\email{sven.auschra@gmail.com}
\uleipT
\author{Viktor Holubec}
\email{viktor.holubec@mff.cuni.cz}
\uleipT
\chau

\date{\today}

\begin{abstract}
Suspensions of motile active particles with space dependent activity form characteristic polarization and density patterns. Recent single-particle studies for planar activity landscapes identified several quantities associated with emergent density-polarization patterns that are solely determined by bulk variables. Naive thermodynamic intuition suggests that these results might hold for arbitrary activity landscapes mediating bulk regions, and thus could be used as benchmarks for simulations and theories.
  However, the considered system operates in a non-equilibrium steady state, and we prove by construction that the quantities in question lose their simple form for curved activity landscapes. Specifically, we provide a detailed analytical study of polarization and density profiles induced by radially symmetric activity steps, and of the total polarization for the case of a general radially symmetric activity landscape. While the qualitative picture is similar to the planar case, all the investigated variables depend not only on bulk variables but also comprise geometry-induced contributions. We verified that all our analytical results agree with exact numerical calculations.
\end{abstract}

\maketitle

\section{Introduction}
\label{sec:introduction}

To feed, hide, or proliferate, both macroscopic ~\cite{Uchida2006,sponberg2017animalLocomotion, Moulia2015} and microscopic~\cite{kaupp2003chemotacticResponse,friedrich2007ChemotaxisSperm,miller2001Quorum,fischer2020QuorumSensing, fischer2020erratum} living organisms
actively adjust their motion to mechanical, optical, or chemical stimuli. The ability to change motility based on the state of the environment is also vital for artificial \emph{motile} active matter ranging from robots~\cite{Mijalkov2016EngineeringBehaviors,Virgh2014} to microscopic active particles \cite{Anderson1989ColloidForces,poon_2013review,zoettl2016EmergBehav,Bechinger2016ActiveEnvironments}, where some of the ultimate goals are noninvasive drug delivery and microsurgeries \cite{patra2013DrugDeliv,Singh2016}. On much lower level of sophistication, large assemblies of active particles exhibit motility-induced phase separation (MIPS) \cite{Cates2015MIPS} into a dense and slow, and dilute and fast phase~\cite{Fily2012AthermalAlignment,Buttinoni2013,redner2013PhaseSep}. Typically but not exclusively, this separation is a consequence of a density dependent propulsion speed \cite{Solon2018GeneralizedMatter,solon2018GenTDofMIPS,bialke2013,speck2015MFT}.

The inhomogenous or space dependent activity comes hand in hand with characteristic modulations in the local density and polarization~\cite{Schnitzer1993TheoryChemotaxis,Sharma2017BrownianActivity,Malakar2018RunAndTumble1D,solon2018GenTDofMIPS,hermann2019MIPS,fischer2020QuorumSensing, fischer2020erratum}. Given the omnipresence of inhomogenous activity at all scales of active matter, the latter can serve as mesoscale indicator for intrinsic microscopic activity in the system~\cite{Soeker2020ActivityFieldsPRL,Auschra2020ActivityFieldsLong}. In spite of that, a thorough investigation of characteristic patterns in the local density and polarization attracted a focused attention of the active matter community only recently. 

It was shown~\cite{Soeker2020ActivityFieldsPRL,Auschra2020ActivityFieldsLong} that density and polarization for a single micrometer-sized Janus swimmer in water are well captured by the active Brownian particle (ABP) model \cite{erdmann2000ABP,schweitzer2007brownianagents,romanczuk2012,cates2013,solon2015} for non-interacting active spheres in a noisy environment. For a single-swimmer and a planar activity interface~\cite{Auschra2020ActivityFieldsLong}, this model allows to identify 
 three quantities that are solely determined by bulk diffusion coefficients, swim speeds, and system size, and thus acquire the status of thermodynamic state variables. Namely (i) the local polarization peak at the interface, (ii) the ratio of densities of the bulk regions on either side of the interface, and (iii) the total polarization caused by the activity step.
 The latter two  maintain this property irrespective of the shape of the (onedimensional) activity modulations, as long as they mediate between two bulk regions \cite{Auschra2020ActivityFieldsLong,hermann2020PolStateFct}. If generally valid, these simple relations can serve as consistency
checks for simulations and benchmarks for theories \cite{hermann2020PolStateFct}.

In this article, we prove by construction that these results in general do not hold for other than planar activity profiles. Concretely, we applied the theoretical framework of Ref.~\cite{Auschra2020ActivityFieldsLong}
to radially symmetric activity steps and investigated in detail the resulting polarization and density patterns. Our analytical results show that the quantities (i)--(iii) depend on the non-zero curvature of the interface and thus on the geometry of the setup.
We also investigate the (radial) total polarization for general radially symmetric motility modulations and show that it acquires a geometry-induced non-local contribution and hence is no longer determined only by bulk variables.
In the limit of vanishing curvature, the obtained results converge to those for planar activity steps~\cite{Auschra2020ActivityFieldsLong,hermann2020PolStateFct}. Our theoretical results can be readily tested using the experimental setup used in Ref.~\cite{soeker2018_thesis}.

Our results would be surprising for a system in thermodynamic equilibrium with solid walls, where their shape does not affect bulk properties. However, they might be expected for the active-matter system at hand, as it operates in a non-equilibrium steady-state. Indeed, the dependence of bulk properties in active-matter systems on the shape of their physical boundaries has been observed in Refs.~\cite{nikola2016curvedWalls,wittmann_2019pressSurfTensCurv,fily2015DensityAtCurvedBoundary,wioland2013confinement}. The dependence on the interface curvature found here can be compared to the Laplace pressure~\cite{butt2013}, e.g. in soap bubbles. The main difference between the two setups is that the increased pressure inside a bubble is caused by a physical force applied in the form of the surface tension by the soap film on the bubble interior. The activity interface in our setup is fixed and the observed influence of its curvature can be traced to geometry-induced imbalance of probability currents across the curved interface.

\section{The Model}
\label{sec:theory}

\begin{figure}[tb!]
  \centering \includegraphics[width=\columnwidth]{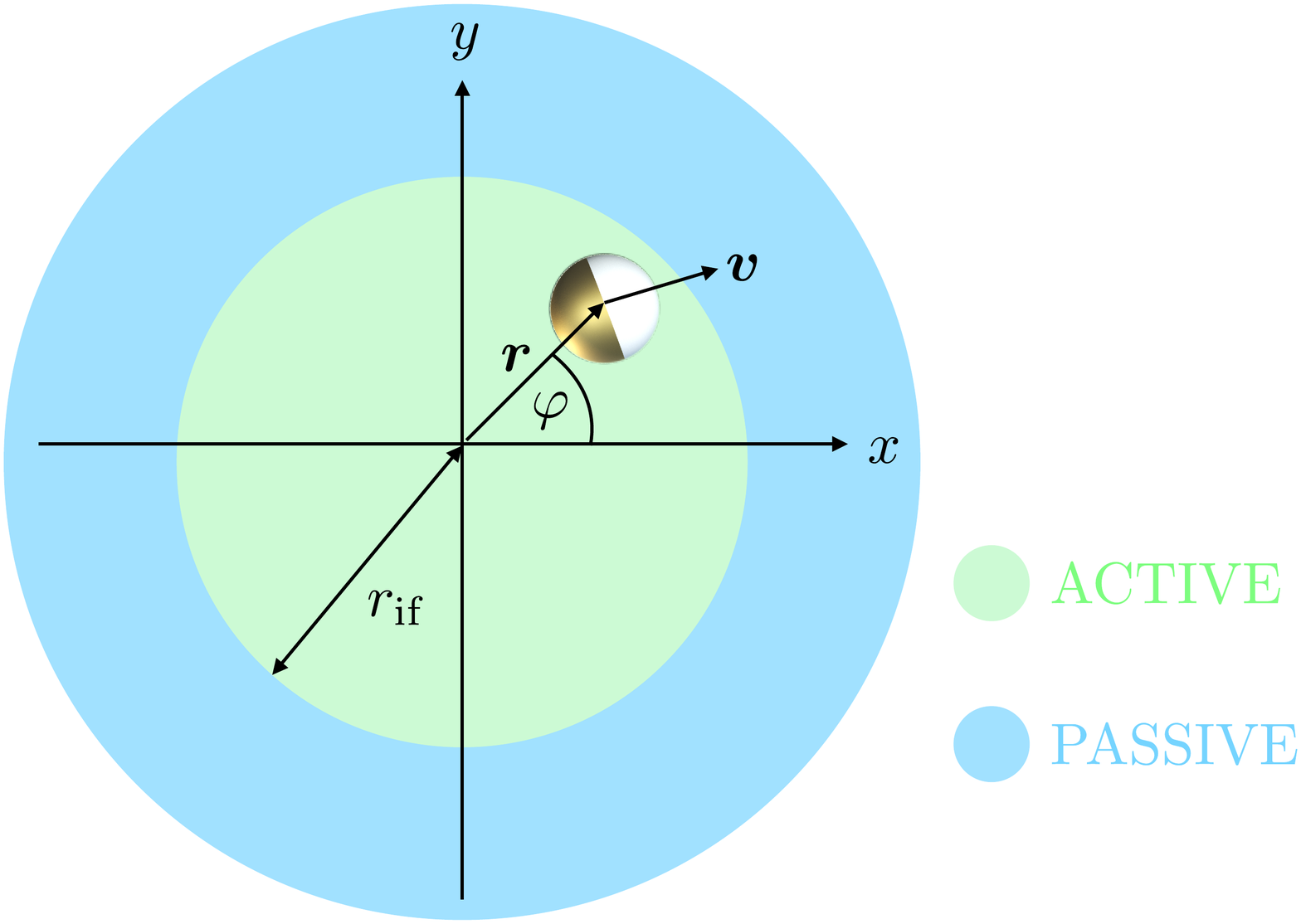}
  \caption{Janus particle with coordinates $x = r \cos\phi$, $y = r \sin\phi$ subjected to a radially symmetric activity profile. The particle propels actively along its orientation $\vv v/v = (\cos\theta, \sin\theta)^\top$ for $r \equiv |\vv r| < \rif$. Otherwise, its swim speed $v$ is zero.}
  \label{fig:setup_radial_theory}
\end{figure}

Consider an overdamped Janus swimmer with space dependent propulsion speed (activity) $v(x,y)$ and orientation parametrized by the angle $\theta$ confined in a plane. For a piecewise constant radially-symmetric activity profile, we depict the system in Fig.~\ref{fig:setup_radial_theory}. We model the particle dynamics by the Active Brownian particle model~\cite{Cates2012DiffusivePhysics} described by the system of Langevin equations
\begin{align}
  \label{eq:langevin_x}
  \partial_t x
  =
  v(x,y) \cos\theta
  +
  \sqrt{2D} \xi_x,
  \\
  \label{eq:langevin_y}
  \partial_t y
  =
  v(x,y) \sin\theta
  +
  \sqrt{2D} \xi_y,
  \\
  \label{eq:langevin_theta}  
  \partial_t \theta
  =
  \sqrt{2\Dr} \xi_\theta.
\end{align}
The transitional and rotational diffusion coefficients $D$ and $\Dr$, respectively,  measure intensities of independent, unit variance, unbiased Gaussian white noise processes $\xi_{x, y, \theta}(t)$.

In the following section, we utilize the framework of Ref.~\cite{Auschra2020ActivityFieldsLong} to derive approximate differential equations for the stationary probability density $\rho(\vv r)$ to find the particle at position $\vv r$ and the corresponding polarization $\vv p(\vv r)$.

\section{Moment Equations}
\label{sec:general}

The dynamic probability density $f(\vv r,\hn,t)$ for finding the Janus swimmer at time $t$ at position $\vv r$ with the orientation $\hn = (\cos\theta, \sin\theta)^\top$, corresponding to the system of stochastic differential equations \eqref{eq:langevin_x}--\eqref{eq:langevin_theta}, obeys the FPE \cite{golestanian2012CollBehav,cates2013,solon2015}
\begin{equation}
  \label{eq:FPE_full}
  \partial_t f
  =
  D \mynabla^2 f
  +
  \Dr \partial_\theta^2 f
  -
  \mynabla \cdot [f v(\vv r) \hn].
\end{equation}
Here, $\partial_t \equiv \partial/\partial t$, and $\mynabla$ represents the Nabla operator with respect to $\vv r$.
The exact moment expansion of $f$ in terms of $\hn$ \cite{bertin2006BoltzmannHydro,golestanian2012CollBehav,cates2013} truncated after the second term reads \cite{Auschra2020ActivityFieldsLong}
\begin{equation}
  \label{eq:moment_expansion_general}
    f(\vv r, \hn, t)
    =
    \frac{1}{2\pi}
    \left[
      \rho(\vv r, t)
      +
      2 \vv p(\vv r, t)
      \cdot \hn
  \right],
\end{equation}
where
\begin{align}
  \label{eq:def_density}
  \rho(\vv r, t)
  &\equiv
    \int \df \hn ~
    f(\vv r, \hn, t),
  \\
  \label{eq:def_polarization}
  \vv p(\vv r, t)
  &\equiv
    \int \df \hn ~
    \hn f(\vv r, \hn, t)
\end{align}
denote time-resolved density and polarization, respectively. 
Multiplying Eq.~\eqref{eq:FPE_full} by 1 or $\hn$, integrating over orientational degrees of freedom, and using the definitions \eqref{eq:def_density} and \eqref{eq:def_polarization},  we obtain the moment equations \cite{golestanian2012CollBehav,Cates2012DiffusivePhysics}
\begin{align}
  \label{eq:mom_equ_rho_general}
  \partial_t \rho(\vv r, t)
  &=
    - \mynabla
    \cdot
    \vv J(\vv r, t),
  \\[0.5em]
  \label{eq:mom_equ_pol_general}
  \partial_t \vv p(\vv r, t)
  &=
    -\Dr \vv p(\vv r, t)
    -
    \mynabla
    \cdot
    \vv M(\vv r, t).
\end{align}
Here, we introduced the (orientation averaged) flux
\begin{equation}
  \label{eq:flux_J_general}
  \vv J(\vv r, t)
  \equiv
  -D \mynabla \rho(\vv r, t)
  +
  v(\vv r)
    \vv p(\vv r, t),
\end{equation}
and the matrix flux
\begin{equation}
  \label{eq:flux_M_general}
  \vv M(\vv r, t)
  \equiv
  -D \mynabla \vv p(\vv r, t)
  +
  \frac{v(\vv r)}{2}
  \rho(\vv r, t)
  \boldsymbol{1},
\end{equation}
with the unit matrix $\boldsymbol{1}$.

Throughout the rest of this article, we will focus on the steady-state solutions $\rho(\vv r)$ and $\vv p(\vv r)$ of Eqs.~\eqref{eq:mom_equ_rho_general} and \eqref{eq:mom_equ_pol_general}, which obey $\partial_t \rho = \partial_t \vv p = 0$ and thus
\begin{align}
  \label{eq:mom_equ_rho_general_steady_state}
  D\mynabla^2 \rho(\vv r)
  &=
    \mynabla \cdot
    \left[
    v(\vv r)
    \vv p(\vv r)    
    \right],
  \\[0.5em]
  \label{eq:mom_equ_pol_general_steady_state}  
  D \mynabla^2 \vv p(\vv r)
  &=
    \Dr \vv p(\vv r)
    +
    \frac12
    \mynabla
    \left[
    v(\vv r)
    \rho(\vv r)
    \right].
\end{align}
Moreover, we assume that, under no-flux boundary conditions, the stationary flux $\vv J(\vv r)$ vanishes. While this  assumption is not generally valid in two (or higher) dimensions, it holds for all setups considered below. Exploiting the no-flux condition in Eq.~\eqref{eq:flux_J_general} and substituting the resulting formula
\begin{equation}
  \label{eq:no_flux_general}
  \mynabla\rho(\vv r)
  =   \frac{v(\vv r)}{D}
  \vv p(\vv r),
  \end{equation}
into Eq.~\eqref{eq:mom_equ_pol_general_steady_state}, we obtain
  \begin{equation}
  \label{eq:mom_equ_pol_plugged_in}  
  \mynabla^2 \vv p(\vv r)
  =     \frac{\vv p(\vv r)}{\lambda^2(\vv r)}
    +
    \frac{\rho(\vv r)}{2D}
    \mynabla
    v(\vv r),
\end{equation}
where we have introduced the length scale
\begin{equation}
  \label{eq:def_lambda_act-pass_circ}
  \lambda(\vv r)
  \equiv
  \left[
    \frac{\Dr}{D}
    +
    \frac{v^2(\vv r)}{2D^2}
  \right]^{-1/2}.
\end{equation}
 A thorough discussion and physical interpretation of this characteristic length scale in the case of a planar motility step is given in Refs.~\cite{Soeker2020ActivityFieldsPRL,Auschra2020ActivityFieldsLong} and we omit it here.

\section{Active-passive interface}
\label{sec:active-pass-interf}

We will now solve Eqs.~\eqref{eq:no_flux_general} and \eqref{eq:mom_equ_pol_plugged_in} for $\rho$ and $\vv p$ for the radially symmetric activity step sketched in Fig.~\ref{fig:setup_radial_theory}. In this setup, the swim speed $v(r) \equiv v_0$, for $r \equiv \sqrt{x^2+y^2} < \rif$, and is zero otherwise.

Polarization and density must reflect the radial symmetry of the activity profile leading to
\(
\rho = \rho(r)
\)
and
\(
\vv p = p(r) \hr,
\)
where $\hr \equiv  \vv r/r =  (\cos\varphi, \sin\varphi)^\top$.
Using this ansatz, the flux-balance condition
\eqref{eq:no_flux_general} 
reduces to
\begin{equation}
  \label{eq:flux_balance_circ}
  \rho'(r)
  =
  \frac{v(r)}{D}p(r),
\end{equation}
where $\rho'(r) \equiv \partial \rho/ \partial r$.
Exploiting this relation in the moment Eq.~\eqref{eq:mom_equ_pol_plugged_in} yields
\begin{equation}
  \label{eq:ode_pol_circ}  
  p''(r)
  =
  -\frac{p'(r)}{r}
  +
  \frac{p(r)}{r^2}
  +
  \frac{p(r)}{\lambda^2(r)}
  +
  \frac{v'(r)\rho(r)}{2D}.
\end{equation}
For the planar setup of Refs.~\cite{Soeker2020ActivityFieldsPRL,Auschra2020ActivityFieldsLong} corresponding to $\rif \to \infty$, the first two terms on the r.h.s.~of this equation are zero, which suggests that the polarization and its derivative decay with the distance from the interface faster than $1/r^2$ and $1/r$, respectively.
In general, the last term of Eq.~\eqref{eq:ode_pol_circ} vanishes everywhere except for $r = \rif$, since $v'(r) = -v_0 \delta(r - \rif)$, with the Dirac delta function $\delta(r)$.
Within the active ($r \leq \rif$) and passive region ($r > \rif$), Eq.~\eqref{eq:ode_pol_circ} reduces to the modified Bessel equation \cite{fischer2020QuorumSensing, fischer2020erratum}. Its general solution reads \cite{abramowitz65_handb}
\begin{equation}
  \label{eq:pol_general_sol_circ}
  p_\te{a,p}(r)
  =
  C_\te{a,p}^{(1)}
  I_1
  \left(
    r / \lambda_\te{a,p}
  \right)
  +
  C_\te{a,p}^{(2)}
  K_1
  \left(
    r / \lambda_\te{a,p}
  \right),
\end{equation}
where $I_m(x)$ and $K_m(x)$ are the modified Bessel functions of the first and second kind, respectively. The characteristic length scales
\begin{equation}
  \label{eq:lambda_a_p}
  \la
  \equiv
  \left(
    \frac{\Dr}{D}
    +
    \frac{v_0^2}{2D^2}
  \right)^{-1/2},
  \qquad
  \lp
  \equiv
  \left(
    \frac{\Dr}{D}
  \right)^{-1/2},
\end{equation}
follow from Eq.~\eqref{eq:def_lambda_act-pass_circ} evaluated in the active and passive region, respectively.

In order to create bulk regions with constant density and vanishing polarization both in the active and in the passive region, we demand in the following that the active-passive interface is far enough both from the origin and from the system's boundary at $r=R$. That is, we assume that $\rif$ and $R-\rif$ are several times greater than $\la$ and $\lp$, respectively. This allows us to apply the boundary conditions
\begin{equation}
  \label{eq:BC_pol}
  \Pa(r = 0) = 0,
  \quad
  \Pp(r = R) = 0.
\end{equation}
Then the general solution \eqref{eq:pol_general_sol_circ} simplifies to
\begin{equation}
  \label{eq:pol_circ_nat_bound}
  p(r)
  =
  \begin{cases}
    \Pa(r) = \Ca I_1(r/\la)
    \quad \hspace{0.2cm} \te{for}
    ~ r \leq \rif
    \\[0.5em]
    \Pp(r) = \Cp K_1(r/\lp)
    \quad \te{for}
    ~ r > \rif
  \end{cases}.
\end{equation}
Integration of Eq.~\eqref{eq:flux_balance_circ} delivers the corresponding density
\begin{equation}
  \label{eq:rho_act_circ_nat_bound}
  \rho(r)
  =
  \begin{cases}
      \Ra
      +
      \frac{v \Ca \la}{D}
      \left[
        I_0(r/\la)-1
      \right]
      \quad \te{for}
      ~r \leq \rif      
      \\[0.5em]
      \Ra
      +
      \frac{v \Ca \la}{D}
      \left[
        I_0(\rif/\la)-1
      \right]
      \equiv \Rp
      \quad \te{for }
      ~r > \rif      
  \end{cases},
\end{equation}
which assumes the bulk value $\Ra \equiv \rho(0)$ in the active and $\Rp \equiv \rho(\rif) \equiv \rho(R)$ in the passive region. The constants $\Ca$ and $\Cp$ in Eq.~\eqref{eq:pol_circ_nat_bound} can be determined from continuity conditions on $p$ and the corresponding flux $\vv M$ at the active-passive interface \cite{risken}. Demanding the polarization $p(r)$ and the projection
\begin{equation}
  \label{eq:M_normal}
  \vv M \cdot \vv{\hat r}
  =
  \left[
    - D p'(r)
    +
    \frac{v(r) \rho(r)}{2}
  \right]
  \vv{\hat r}
\end{equation}
of the matrix flux \eqref{eq:flux_M_general} onto the radial direction to be continuous at $r = \rif$ renders
\begin{align}
  \label{eq:matching_intuit}
  \Pa(\rif)
  &=
    \Pp(\rif),
  \\[0.5em]
  \label{eq:jump_condition}
  \Pa'(\rif) - \Pp'(\rif)
  &=
    \frac{v_0}{2D} \rho(\rif).
\end{align}
The density \eqref{eq:rho_act_circ_nat_bound} satisfies $\Ra(\rif) = \Rp(\rif)$ by construction. The normal component $\vv J \cdot \vv{\hat r}$ is continuous due to the imposed no-flux condition $\vv J \equiv \vv 0$.
The constant $\Ra$ in Eq.~\eqref{eq:rho_act_circ_nat_bound} follows from the normalization condition
\begin{equation}
  \label{eq:normalization}
  \int\limits_0^{2\pi} \df \phi
  \int\limits_0^R\df r ~ r \rho(r)
  =
  1.
\end{equation}
\begin{figure*}[tb!]
  \centering
  \includegraphics[width=\linewidth]{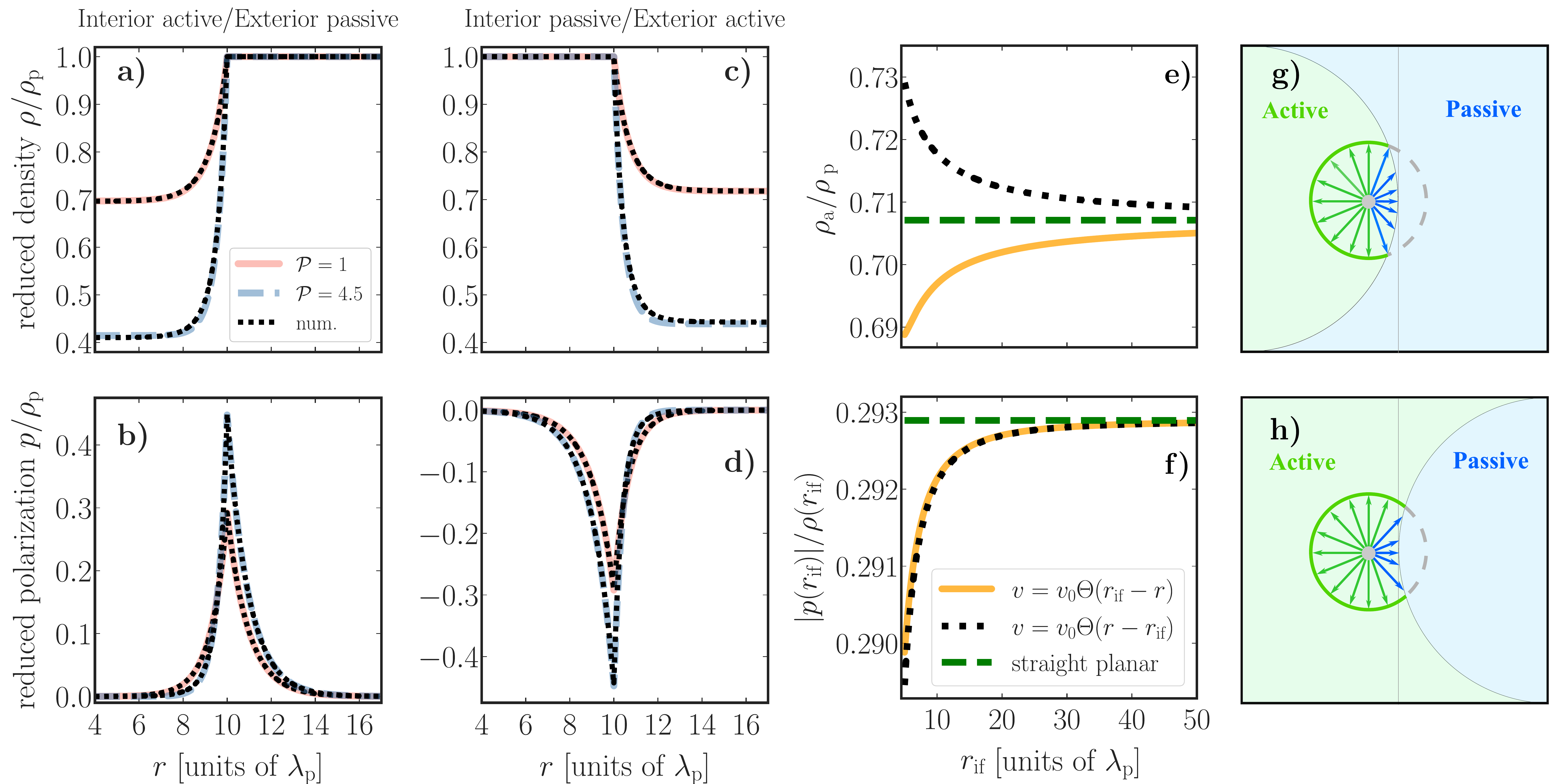}
  \caption{\textbf{a)-d):} Reduced density (top) and polarization (bottom) profiles near a radially symmetric active-passive [a),b)] / passive-active [c)/d)] interface at $r = \rif$. Lengths are measured in units of $\lp=\sqrt{D/\Dr}$ and velocities in units of $\sqrt{2 D \Dr}$. This corresponds to a dimensionless theoretical description in terms of the P\'{e}clet number~$\mathcal P$ \eqref{eq:Peclet}. Theory profiles (solid/dashed curves) were calculated for two distinct $\mathcal P$ and $\rif = \SI{10}{\lp}$ using Eqs.~\eqref{eq:pol_circ_nat_bound}-\eqref{eq:normalization} in a) and b) and Eqs.~\eqref{eq:polarization_active_circ}-\eqref{eq:Ca_2} in c) and d), and compared against exact numerically obtained~\cite{holubec2018} profiles (dotted lines). \textbf{e), f)}: Density ratio, $\Ra/\Rp$, and magnitude of the (reduced) polarization at the interface, $|p(\rif)|/\rho(\rif)$, for $\mathcal P = 1$ as functions of the radial distance $\rif$ of the interface. Plotted curves correspond to the following analytical expressions (setups): Solid curves (interior active/exterior passive): Eqs.~\eqref{eq:Pmax_circ} and \eqref{eq:density_ratio_circ}; Dotted curves (interior passive/exterior active): Eqs.~\eqref{eq:Pmax_circ_p} and \eqref{eq:density_ratio_circ_p}; Dashed lines (straight planar case): Eqs.~\eqref{eq:Pmax_limit} and \eqref{eq:density_ratio_limit}. \textbf{g)/ h)}: Particle inside the active region in the vicinity of a concave/convex active-passive interface. In both panels, the vertical line corresponds to a straight active-passive interface. Relative to the latter case, for a concave/convex geometry, the particle has a higher/lower chance to end up in the passive region.
    }
  \label{fig:pol_rho_th_vs_num}
\end{figure*}

Figures~\ref{fig:pol_rho_th_vs_num} a) and b) show nice agreement of the approximate analytic density and polarization profiles \eqref{eq:pol_circ_nat_bound} and \eqref{eq:rho_act_circ_nat_bound} with exact numerical solutions \cite{holubec2018} for two distinct particle activities, expressed in terms of the P\'{e}clet number
\begin{equation}
  \label{eq:Peclet}
  \mathcal P
  \equiv
  \frac{v_0^2}{2 D \Dr}.
\end{equation}
We observe nice agreement with exact numerical solutions \cite{holubec2018} in both cases.  
In Sec.~\ref{sec:appr-glob-sum}, we show that the approximate solutions deviate from the exact results for much smaller P\'{e}clet numbers than $\mathcal P = 100$ found for straight planar interfaces~\cite{rodenburg2020_thesis}. This is because Eq.~\eqref{eq:ode_pol_circ} cannot be mapped onto an exactly solvable (2-species) run-and-tumble model \cite{Malakar2018RunAndTumble1D} providing the same phenomenology as the full model\,\footnote{%
  In the case of planar activity steps, the corresponding approximate moment equations, similar to Eqs.~\eqref{eq:flux_balance_circ} and \eqref{eq:ode_pol_circ}, could be mapped onto an exact one-dimensional run-and-tumble model of left or right-moving particles~\cite{Auschra2020ActivityFieldsLong}. This 2-species model robustly captured the essential physics behind the formal results obtained for planar steps and justifies the large range of P\'{e}clet numbers over which the approximate moment equations [cf.~Eqs.~\eqref{eq:flux_balance_circ} and \eqref{eq:ode_pol_circ}] deliver accurate results. Since this equivalence between the two models is no longer present, deviations between approximate analytic and exact numerical results already occur for smaller P\'{e}clet numbers.}, which was the key ingredient for success of the approximate solutions at planar interfaces~\cite{Soeker2020ActivityFieldsPRL,Auschra2020ActivityFieldsLong}.

The qualitative behavior of density and polarization profiles is the same as for a planar activity step \cite{Soeker2020ActivityFieldsPRL,Auschra2020ActivityFieldsLong}. Namely, an increased activity step induces a higher polarization and a larger ratio $\rho_\te{p}/\rho_\te{a}$ of bulk densities of the passive and active regions. The polarization peaks exactly at the active-passive interface and decays over characteristic lengths $\lp$ and $\la < \lp$ into the passive and active region, respectively.  The density profile remains constant at the bulk density $\rho = \Rp$ throughout the whole passive region. On crossing the interface, it decays to the bulk density $\Ra < \Rp$ pertaining to the active region over a length scale $\la$. We refer to Refs.~\cite{Soeker2020ActivityFieldsPRL,Auschra2020ActivityFieldsLong} for a more detailed physical interpretation and discussion of the emerging polarization and density variations. Here, we focus on the influence of curvature on these profiles.

In Figs.~\ref{fig:pol_rho_th_vs_num} c) and d), we show the (reduced) density and polarization profiles corresponding to the inverse setup for which the particle is passive for $r < \rif$ and active otherwise. The derivation of the analytic profiles (solid/dashed lines) is similar to the above calculations, and is detailed in App.~\ref{sec:reversed-situation}. The approximate theory profiles overlap with the corresponding exact numerical solutions. The qualitative picture is similar to the situation shown in Figs.~\ref{fig:pol_rho_th_vs_num} a) and b), with flipped active/passive regions, and a negative polarization in the vicinity of the interface, confirming that the particle preferably points into the passive region \cite{Soeker2020ActivityFieldsPRL,Auschra2020ActivityFieldsLong}. Note, however, that the convexity/concavity of the activity interface also leads to quantitative differences between the two cases.

To grasp the influence of the curvature of the activity step more quantitatively, we compare the maximum relative polarization, $p(\rif)/\rho(\rif)$, which constitutes a suitable order parameter for the polarization at the interface, and the bulk density ratio, $\Ra/\Rp$, for circular and straight planar interfaces.
For the setup where the particle is active for $r<\rif$ and passive otherwise [Fig.~\ref{fig:pol_rho_th_vs_num} a),b)], these quantities are given by
\begin{align}
  \label{eq:Pmax_circ}
  \frac{p(\rif)}{\rho(\rif)}
  &=
    \frac{v_0}{2D}
    G_\te{A}(\rif),
  \\[0.5em]
  \label{eq:density_ratio_circ}
  \frac{\Ra}{\Rp}
  &=
    1
    -
    \frac{v_0^2 \la}{2 D^2}
    G_\te{A}(\rif)
    \left(
    \frac{I_0}{I_1}
    -
    \frac{1}{I_1}
    \right),
\end{align}
as detailed in App.~\ref{sec:circular-setup}.
Here, the (geometry) function reads
\begin{equation}
  \label{eq:geometry_fct}
  G_\te{A}(\rif)
  \equiv
  \left(  
    \frac{I_0+I_2}{2\la I_1}
    +
    \frac{K_0+K_2}{2\lp K_1}    
  \right)^{-1},
\end{equation}
with $I_m \equiv I_m(\rif/\la)$ and $K_m \equiv K_m(\rif/\lp)$. The subscript ``A'' indicates that $G_\te{A}$ corresponds to the case where the particle is \emph{active} for $r<\rif$. The geometry function $G_\te{P}(\rif)$ for the inverted setup is derived in App.~\ref{sec:reversed-situation}.  Exploiting the asymptotic expansions
\(
I_n(z)
\sim
\te{e}^{z}/\sqrt{2 \pi z}
\)
and
\(
K_n(z)
\sim
\te{e}^{-z}/\sqrt{2z / \pi}
\),
valid for $z \gg 1$ irrespective of the order $n$ \cite{abramowitz65_handb},
one finds
\begin{equation}
  \label{eq:G_limit}
  G_\te{A}(\rif)
  \sim
  \frac{\la \lp}{\la + \lp}.
\end{equation}

For $\rif \gg \lp > \la$, 
the maximum (relative) polarization therefore approaches
\begin{equation}
  \label{eq:Pmax_limit}
  \frac{p(\rif)}{\rho(\rif)}
  \sim
  \frac{v_0}{2D}\frac{\la \lp}{\la + \lp}
  =
  \frac{1}{\sqrt{2}}
  \frac{\sqrt{\mathcal P}}{1+\sqrt{1+\mathcal P}},
\end{equation}
which coincides with the expression found for planar interfaces \cite{Auschra2020ActivityFieldsLong}. The corresponding asymptotic behavior of the density ratio~\eqref{eq:density_ratio_circ}
\begin{equation}
  \label{eq:density_ratio_asymp}  
  \frac{\Ra}{\Rp}
  \sim 
 1-
  \frac{v_0^2\la}{2D^2}
  \frac{\la \lp}{\la + \lp}
    \left(
    1 - \sqrt{ \frac{2 \pi \rif}{\la} }
    \te{e}^{-\rif/\la}
    \right)
  \end{equation}
  still displays an exponential decaying with $\rif$. By taking the limit $\rif \to \infty$, it reduces to the result found at planar activity steps \cite{Auschra2020ActivityFieldsLong}
  \begin{equation}
    \label{eq:density_ratio_limit}
    \frac{\Ra}{\Rp}
    =
    \frac{\la}{\lp}
    =
    \frac{1}{\sqrt{1 + \mathcal P}}.
  \end{equation}
   The analytic expressions for $p(\rif)/\rho(\rif)$ and $\Ra/\Rp$ for the inverted setup [Fig.~\ref{fig:pol_rho_th_vs_num} c),d)] are derived in App.~\ref{sec:reversed-situation}  along similar lines. 

    Figures~\ref{fig:pol_rho_th_vs_num} e) and f) show the dependence of these quantities on the radius of curvature $\rif$ of the interface for both circular setups, as well as their their counterparts~\eqref{eq:Pmax_limit} and \eqref{eq:density_ratio_limit} for straight planar interfaces (dashed horizontal lines).
    With increasing $\rif$, the polarization peaks $|p(\rif)|/\rho(\rif)$ for the circular setups approach the one for the straight motility step from below. For the setup where the particle is active for $r < \rif$ (solid curve), the peak is slightly larger ($\approx \SI{0.02}{\percent}$ for $\rif = \SI{10}{\lp}$) than for the inverted setup (dotted curve). Turning to the bulk density ratio, if the interior is active (solid line), the bulk density ratio $\Ra/\Rp$ is smaller as compared to the straight planar case, whereas it exceeds it for the inverse circular setup (dotted curve).
    
    This behavior can be intuitively understood as a result of a geometry-induced imbalance in probability fluxes across the curved interface. Consider the situations sketched in Fig.~\ref{fig:pol_rho_th_vs_num} g) and h). In both panels, a Janus particle (gray dot) is situated inside an active region, in a close proximity to an adjacent passive region. In g), the active-passive interface is concave, whereas in h), it is convex. In both panels, the vertical lines correspond to a straight active-passive interface. For simplicity, consider a quasi-ballistic particle motion denoted by the arrows in both panels. As indicated by the number of blue arrows relative to the green ones, the particle's chance to enter the passive region is higher for the concave [g)]  than convex [h)] geometry. The setup with active interior [g)] thus yields a larger bulk density ratio $\Ra/\Rp$ than the inverse setup [ h)]. It follows that the density ratio corresponding to a concave/convex active-passive interface is always smaller/larger than its counterpart for a straight planar interface. As the curvature of the circular activity interface decreases, \ie~for $\rif \to \infty$, the bulk density ratio for straight interfaces is approached.

To gain an intuition on why the magnitude of the reduced polarization, $|p(\rif)|/\rho(\rif)$, is  always larger in the straight planar case than for a circular interface [see Fig.~\ref{fig:pol_rho_th_vs_num} f)] is more difficult.
    The absolute value of the polarization $|p(\rif)|$ depends on the probability that the particle with a given orientation hits the interface. Compared to the planar case, for the concave interface shown in g), there are less active particles in the bulk to hit the interface with a broader range of polarizations, and vice versa for the convex interface shown in h). Hence we observe three competing ingredients that determine the absolute polarization in the concave (convex) case: low (high) bulk density in the active region, large (small) probability for a given particle to hit the interface, and large (small) average polarization of particles which hit the interface, where the strength of the individual ingredients is compared to the planar case. Furthermore, the magnitude of the reduced polarization is obtained as absolute polarization divided by density of the passive bulk, which is high for the concave and low for the convex case. We thus find in both circular setups two ingredients leading to an increase and two leading to a decrease of $|p(\rif)|/\rho(\rif)$. Our analytical results show that they compensate each other in such a way that the magnitude of the reduced polarization for circular interfaces is always lower than in the planar case. Unfortunately, it seems impossible to guess the influence of these ingredients based on physical intuition. 

The maximum polarization and density ratios, which are for planar motility steps solely determined by bulk variables $(v_0, D, \Dr)$~\cite{Soeker2020ActivityFieldsPRL,Auschra2020ActivityFieldsLong} thus depend on the interface radius $\rif$ in case of a circular activity step. 
This suggests that arbitrarily curved activity steps generally yield geometry-induced contributions to the emergent density-polarization patterns, and the corresponding maximum polarization and bulk density ratio. To provide further evidence for this conjecture, we now study the total polarization, which is also solely determined by bulk quantities in the case of planar interfaces~\cite{hermann2020PolStateFct}, for arbitrary radially symmetric activity modulations.
\section{Total polarization}
\label{sec:total-polarization}

Without alignment forces, local polarization in active-matter systems arises from spacial sorting of particles with different orientations. Therefore the total polarization vector, $\boldsymbol{\mathcal P}_\te{tot}$, defined as the integral
\begin{equation}
  \label{eq:P_tot_general}
  \boldsymbol{\mathcal P}_\te{tot}
  \equiv
  \int_{\mathcal V} \df \vv r ~ \vv p(\vv r)
\end{equation}
over the whole space $\mathcal V$, universally vanishes for systems with no-flux boundary conditions \cite{hermann2020PolStateFct}. 

A more appropriate definition of total polarization induced by activity landscapes that mediate between two bulk regions is to restrict the domain of integration $\mathcal V$ so that it connects the two bulk regions.
For planar activity profiles, it is natural to integrate along a ray of fixed width parallel to the x-axis and thus perpendicular to the interface. Then, the magnitude $P_\te{tot}$ of such defined total polarization is proportional to the difference in strengths of fluxes, $v \rho$, corresponding to the two bulk regions \cite{Auschra2020ActivityFieldsLong,Soeker2020ActivityFieldsPRL}, and thus it acquires the status of a thermodynamic state variable. 
For an arbitrary radial activity profile $v(r)$ that mediates between two bulk regions of respectively constant activity, the radial symmetry implies that the local polarization profile must be of the form
\(
\vv p = p(r) \vv {\hat r},
\)
with
\(
\vv {\hat r} = (\cos\phi, \sin\phi)^\top
\).
To, match the planar definition, we define the magnitude of the total radial polarization as the integral
\begin{equation}
  \label{eq:P_tot_radial}
  P_\te{tot}
  \equiv
  \int\limits_{R_1}^{R_2} \df r ~ 
  p(r),
\end{equation}
of the projection $p(r)$ of the polarization vector onto the radial axis over a ray of fixed infinitesimal width, perpendicular to the interface, and mediating the inner bulk region at radius $R_1$ and the outer one at $R_2$
 (see Fig.~\ref{fig:setup_psi}).
Alternatively, and more naturally from the point of view of polar coordinates, one could integrate the polarization over an infinitesimal wedge mediating the two bulks. This would correspond to substituting $r p(r)$ for $p(r)$ in the definition~\eqref{eq:P_tot_radial}. However, in this case, the width of the integration region increases with $r$.

Below we show that $P_\te{tot}$ is generally composed of a contribution proportional to the difference
\(
v(R_1)\rho(R_1) - v(R_2)\rho(R_2)
\)
of the flux strengths, as for planar interfaces \cite{hermann2020PolStateFct}, and a second non-local contribution induced by the non-zero curvature of the interface. A similar expression also holds for the alternative definition of $P_{\te tot}$ with $r p(r)$.

\subsection{Derivation of total polarization}
\label{sec:derivation}

\begin{figure}[tb!]
  \centering
  \includegraphics[width=\columnwidth]{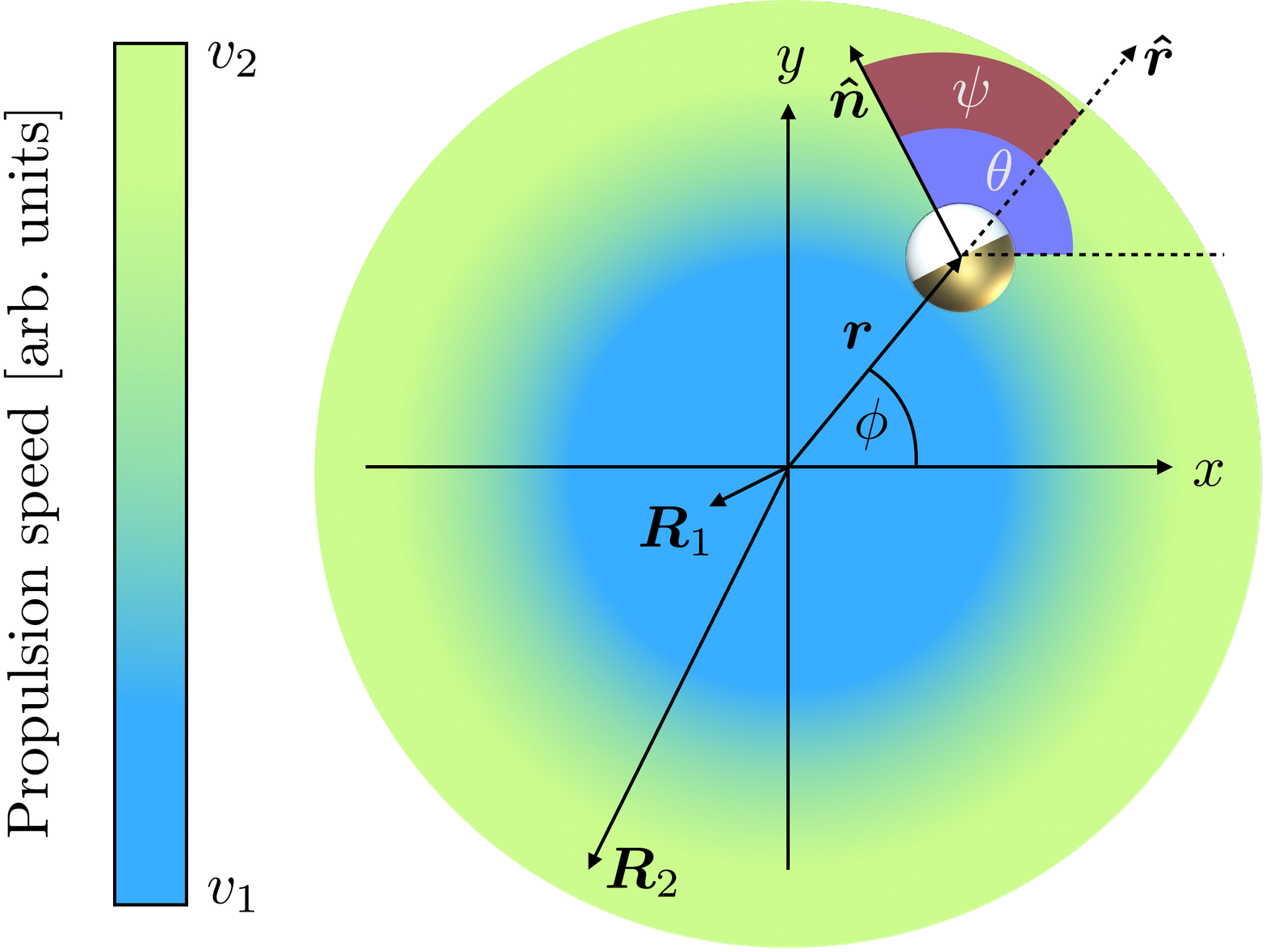}
  \caption{A Janus particle at position
    \(
    \vv r = (r \cos \phi, r \sin\phi)^\top
    \)
    actively propelling along its orientation
    \(
    \vv {\hat n} = (\cos \theta, \sin \theta)^\top
    \).
    The particle's orientation relative to the radial unit vector $\vv{\hat r}$ is measured by the angle $\psi$.    
    The swimmer's propulsion speed follows a radially symmetric activity profile (color-coded), which mediates between an inner (dark blue, $\vv R_1$) and outer (light green, $\vv R_2$) bulk region with different constant activities. 
  }
  \label{fig:setup_psi}
\end{figure}

We introduce polar coordinates, $x = r \cos \phi$ and $y = r \sin\phi$, and the angular variable $\psi \equiv \theta - \phi$, which measures the particle orientation relative to the radial axis (see Fig.~\ref{fig:setup_psi}). Under this transformation, the system of three Langevin equations \eqref{eq:langevin_x}-\eqref{eq:langevin_theta} reduces to the two-dimensional set
\begin{align}
  \label{eq:langevin_r}
  \partial_t r
  &=
    v(r) \cos\psi
    +
    \sqrt{2 D} \xi_r,
  \\[0.5em]
  \partial_t \psi
  &=
    -\frac{v(r)}{r} \sin\psi
    +
    \sqrt{ 2
    \left(
    \frac{D}{r^2}
    +
    \Dr
    \right) }
    \xi_\psi,
\end{align}
where $\xi_r$ and $\xi_\psi$ denote independent, zero-mean, unbiased Gaussian white noise processes.
The associated FPE for the stationary probability density $\frak f(r,\psi)$ for finding the particle at distance $r$ with relative orientation $\psi$ reads
\begin{equation}
  \label{eq:FPE_transf}
  0
  =
  - \partial_r \frak J
  +
  \left(
    \frac{D}{r^2}
    +
    \Dr
  \right)
  \partial_\psi^2 \frak f
  +
  \frac{v(r)}{r}
  \partial_\psi
  \left(
    \sin \psi \frak f
  \right),
\end{equation}
where we have introduced the (angle-resolved) flux
\begin{equation}
  \label{eq:J_transf}
  \frak J(r, \psi)
  \equiv
  -D \partial_r \frak f
  +
  \frac{D}{r} \frak f
  +
  \cos\psi ~ v(r) \frak f.
\end{equation}
Due to the radial symmetry, $\frak f$ and $\frak J$ must be even functions of $\psi$, \ie,
\(
\frak f(r,\psi) = \frak f(r,-\psi),
\)
and
\(
\frak J(r,\psi) = \frak J(r,-\psi).
\)
The Fourier expansion of the flux $\frak J$ thus reads
\begin{equation}
  \label{eq:J_fourier}
  \frak J(r,\psi)
  =
  \sum\limits_{n=0}^\infty \frak J_n(r) \cos(n\psi),
\end{equation}
where $\frak J_n(r)$ is the $n$th Fourier coefficient. Plugging this series into the FPE \eqref{eq:FPE_transf} and integrating twice over the angle from $0$ to $\psi$ yields
\begin{equation}
  \label{eq:density_flux_relation}
  \begin{split}
    \left(
      \frac{D}{r^2}
      +
      \Dr
    \right)
    \frak f(r,\psi)
    =~
    &\frak f_0(r)
    -
    \sum\limits_{n=1}^\infty
    \partial_r \frak J_n(r) \frac{\cos(n\psi)}{n^2}
    \\
    &-
    \frac{v(r)}{r}
    \int\limits_0^{\psi} \df \tilde \psi ~
    \sin\tilde\psi ~ \frak f(r,\tilde \psi).
  \end{split}
\end{equation}
The unknown function $\frak f_0(r)$ stems from the second integration. The integration constant from the first integration renders, after the second integration, a term linear in $\psi$, and thus it must be zero to maintain periodicity.

The radial component of the local polarization vector $\boldsymbol{\frak p}(r)$ is defined by
\begin{equation}
  \label{eq:pol_transf}
  \frak p(r)
  \equiv
  \int\limits_0^{2 \pi} \df \psi
  \cos\psi ~ \frak f(r,\psi).
\end{equation}
The corresponding tangential component
\(
  \int_0^{2 \pi} \df \psi
  \sin\psi ~ \frak f(r,\psi)
  \)
vanishes due to the radial symmetry.
 Multiplying Eq.~\eqref{eq:density_flux_relation} by $\cos \psi$, integrating over $\psi$ from 0 to $2\pi$, and using orthogonality of trigonometric functions renders
\begin{equation}
  \label{eq:local_sum_rule}
  \left(
    \frac{D}{r^2}
    +
    \Dr
  \right)
  \frak p(r)
  =
  -\pi \partial_r\frak J_1(r)
  +
  \frac{v(r)}{r}
  \int\limits_0^{2\pi}
  \df \psi ~ \sin^2\psi ~ \frak f(r, \psi).
\end{equation}
The integral on the r.h.s. was obtained by interchanging the order of the double integration\,\footnote{
  \[
    \int\limits_0^{2\pi} \df \psi ~
    \cos \psi
    \int\limits_0^\psi \df \tilde \psi ~
    \frak f(r, \tilde \psi) \sin \tilde \psi
    =
    \int\limits_0^{2\pi} \df \tilde \psi ~
    \frak f(r, \tilde \psi) \sin \tilde \psi
    \int\limits_{\tilde \psi}^{2\pi} \df \psi ~
    \cos \psi
  \]
}.
Using the definitions \eqref{eq:J_transf} and \eqref{eq:pol_transf}, we find that the first coefficient $\frak J_1$ of the Fourier series \eqref{eq:J_fourier} in terms of $\frak f$ and $\frak p$ reads
\begin{align}
  \frak J_1
  &=
    \frac{1}{\pi}
    \int\limits_0^{2\pi} \df \psi ~
    \frak J(r,\psi) \cos\psi
  \\[0.5em]
  \label{eq:J1}
  &=
    \frac{1}{\pi}
    \left( 
    \frac{D}{r} \frak p
    -
    D \partial_r \frak p
    \right)
    +
    \frac{v}{\pi}
    \int\limits_0^{2\pi} \df \psi ~
    \frak f(r,\psi) \cos^2\psi.
\end{align}

The distributions $\frak f$ and $f$ and the corresponding polarizations $\frak p$ and $p$
are connected via the Jacobian \cite{risken}
\(
\left|
  \partial(x,y)/\partial(r,\phi)
\right|
=
r
\)
, \ie, $  \frak f = r f$ and
$  \frak p = r p$.
Plugging these transformations and Eq.~\eqref{eq:J1} into Eq.~\eqref{eq:local_sum_rule} yields
\begin{align}
  \Dr p
  =
    &D \partial_r
    \left(
    \frac{p}{r} + \partial_r p
    \right)  
  -\partial_r
    \left(
    v \langle \cos^2\psi \rangle
    \right)
  -\frac{v}{r}
    \langle 2 \cos^2\psi - 1 \rangle.
    \label{eq:derivXX}
\end{align}
Here, the averaging is defined as
\begin{equation}
  \label{eq:def_average}
  \langle \bullet \rangle
  \equiv
  \int\limits_0^{2\pi} \df \psi ~
  \bullet f(r, \psi).
\end{equation}
Finally, Eqs.~\eqref{eq:P_tot_radial} and
\(
2 \cos^2\psi -1 = \cos(2\psi)
\)
render the closed expression for the total polarization:
\begin{equation}
  \label{eq:P_tot_res_general}
  P_\te{tot}
  =
  \frac{D}{\Dr}
  \left.
    \left(
    \frac{p}{r} + \partial_r p
    \right)    
  \right|_{R_1}^{R_2}  
  \left.
    -\frac{v}{\Dr}
    \langle \cos^2\psi \rangle
  \right|_{R_1}^{R_2}
  -
  \mathcal I[v](R_1,R_2),
\end{equation}
where we introduced the functional
\begin{equation}
  \label{eq:P_tot_nonlocal}
  \mathcal I[v](R_1,R_2)
  \equiv
  \int\limits_{R_1}^{R_2} \df r ~
  \frac{v(r)}{r \Dr}
  \langle \cos(2\psi) \rangle.
\end{equation}
Within bulk regions, we have
\(
p(R_{1/2}) = \partial_r p(R_{1/2}) = 0
\)
and
\(
f(R_{1/2},\psi) = \rho(R_{1/2})/(2\pi)
\). Hence, the first term on the r.h.s.~of Eq.~\eqref{eq:P_tot_res_general} vanishes and the second one simplifies to
\(
\langle \cos^2 \psi \rangle(R_{1/2})
=
\rho(R_{1/2})/2.
\)
The total polarization between two bulk regions thus reads
\begin{equation}
  \label{eq:P_tot_final}
    P_\te{tot}
    =
    \frac{
      v(R_1) \rho(R_1) - v(R_2) \rho(R_2)
    }{2 \Dr}
    -
    \mathcal I[v](R).
  \end{equation}
The first summand above coincides with the total polarization found for planar interfaces \cite{Soeker2020ActivityFieldsPRL,Auschra2020ActivityFieldsLong}. The second term is a non-local contribution attributed to the non-zero curvature of the considered activity profile. It vanishes when the activity profile becomes effectively planar, i.e. when its radius diverges while the  thickness measured by the distance between the two bulk regions remains finite.
The (radial) total polarization for curved activity profiles is thus not solely determined by stationary properties of the bulk, and, in this sense, looses its status of a state variable.
Using Eq.~\eqref{eq:local_sum_rule}, one can show along similar lines as above that a similar expression would be obtained for the alternative definition of the total polarization using $\frak p$ instead of $p$ in Eq.~\eqref{eq:P_tot_radial}. Specifically, the first term in this total polarization follows from~Eq.~\eqref{eq:P_tot_final} after substituting $r \rho$ for $\rho$ in the first term on the r.h.s. The curvature dependent term changes more but also vanishes for diverging radius of the interface.

 We now demonstrate that $\mathcal I[v](R_1, R_2)$ in Eq.~\eqref{eq:P_tot_final} vanishes when truncating the exact moment expansion of $f(r,\psi)$ after two terms as in Eq.~\eqref{eq:moment_expansion_general} and quantify deviations between the exact solution and approximate solution of Sec.~\ref{sec:general}.
  
\subsection{Approximate global sum rule and deviations}
\label{sec:appr-glob-sum}

Plugging $\rho=\rho(r)$ and $\vv p = p(r) \hat{\vv r}$ in to the truncated moment expansion \eqref{eq:moment_expansion_general}, the approximate distribution function can be written as
\begin{equation}
  \label{eq:f_approx_psi}
  f(r,\psi)
  =
  \frac{1}{2\pi}
  \left[
    \rho(r)
    +
    p(r) \cos\psi
  \right].
\end{equation}
Using the average~\eqref{eq:def_average}, the density $\rho(r)$ and polarization $p(r)$ are given by $\langle 1 \rangle$ and $\langle \cos\psi \rangle$, respectively.
All higher (angular) moments within the approximation~\eqref{eq:f_approx_psi} vanish. In particular,
\(
\langle \cos(2\psi) \rangle = 0
\)
and the functional $\mathcal I$~\eqref{eq:P_tot_nonlocal} gives zero. The total polarization \eqref{eq:P_tot_final} then reads
\begin{equation}
  \label{eq:P_tot_approx}
    P_\te{tot}
    =
    \frac{
      v(R_1) \rho(R_1) - v(R_2) \rho(R_2)
    }{2 \Dr},
  \end{equation}
  which is the result found for onedimensional activity landscapes \cite{Auschra2020ActivityFieldsLong,hermann2020PolStateFct}. For the radial activity profiles considered here, it only holds within the approximation $\langle \cos(n\psi) \rangle=0$ for $n>1$.

To verify the exact analytical result~\eqref{eq:P_tot_final} for the total polarization, and to assess the scope of the approximation \eqref{eq:P_tot_approx}, we numerically calculated the exact distributions $f(r,\psi)$ \cite{holubec2018} for several radially symmetric activity steps. As in Sec.~\ref{sec:active-pass-interf}, the particle propels actively if its radial distance $r<\rif$ and its swimming mechanism is switched off otherwise (see Fig.~\ref{fig:setup_radial_theory}).
\begin{figure}[tb!]
  \centering
  \includegraphics[width=\columnwidth]{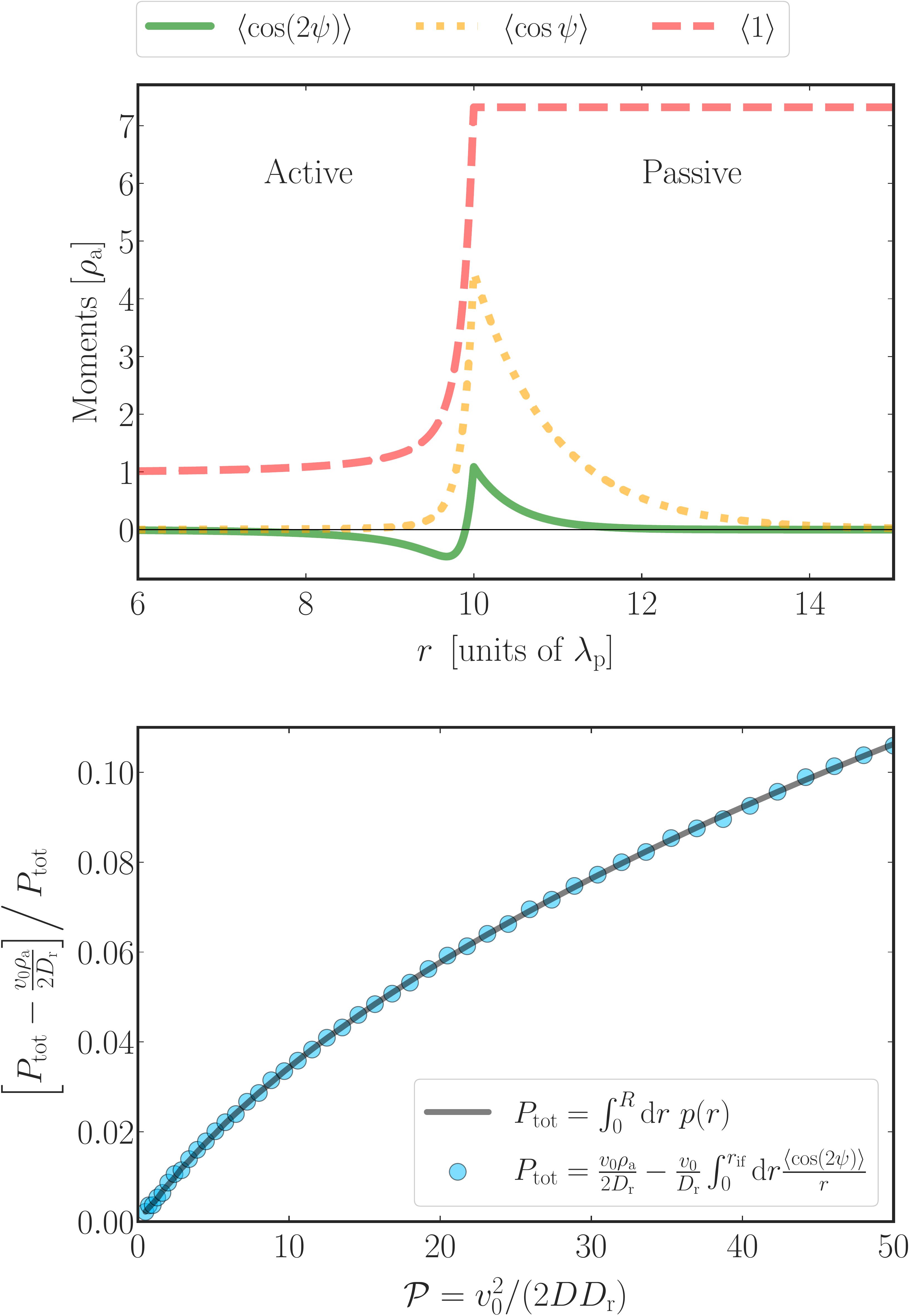}
  \caption{Moments (Eq.~\eqref{eq:def_average}, top) and relative deviations of the total polarization (Eq.~\eqref{eq:P_tot_radial}, bottom) from its approximate value \eqref{eq:P_tot_approx} in the vicinity of a radially symmetric active-passive interface at $r = \rif = \SI{10}{\lp}$. The particle propels actively if its radial distance $r<\rif$ and only diffuses otherwise. Lengths are measured in units of $\lp=\sqrt{D/\Dr}$ and velocities in units of $\sqrt{2 D \Dr}$. The overall radial extension of the system was chosen $R=\SI{20}{\lp}$, and in the top figure we took the P\'{e}clet number $\mathcal P = v_0/(2 D \Dr) = 40$. The presented data was obtained using the exact numerically determined distribution $f(r,\psi)$ \cite{holubec2018}. The solid line and circles in the bottom panel were determined from Eqs.~\eqref{eq:P_tot_radial} and \eqref{eq:P_tot_final}, respectively.}
  \label{fig:total_polarization}
\end{figure}

The upper panel of Fig.~\ref{fig:total_polarization} depicts the moments $\left<\cos (n \psi) \right>$, $n=0,1,2$ calculated from Eq.~\eqref{eq:def_average} using the exact distribution $f(r,\psi)$ for  P\'{e}clet number $\mathcal P = 40$. The moments are normalized by the bulk density $\Ra$ in the active region. In the active region, the second moment $\langle \cos(2\psi) \rangle$ is up to the active-passive interface largely negative rendering its contribution \eqref{eq:P_tot_nonlocal} to the total polarization~\eqref{eq:P_tot_final} positive. 
Since the amplitude of the second moment increases with P\'{e}clet number, the approximate result for the total polarization~\eqref{eq:P_tot_approx} underestimates its exact value the more the larger the particle activity, 
as can be inferred from the lower panel of Fig.~\ref{fig:total_polarization}.
It shows that the relative deviation between the two reaches roughly $\SI{10}{\percent}$ for $\mathcal P = 50$ and thus the approximation~\eqref{eq:P_tot_approx} is reasonably accurate for experimentally realizable P\'{e}clet numbers~\cite{Soeker2020ActivityFieldsPRL}.
The lower panel of Fig.~\ref{fig:total_polarization} also shows perfect agreement between the analytical result~\eqref{eq:P_tot_final} and polarization evaluated form Eq.~\eqref{eq:P_tot_radial} using the numerically determined density $f(r,\psi)$. 
  
 \section{Conclusion}
\label{sec:conclusion}
We derived approximate analytical formulas for polarization and density profiles induced by a radially symmetric motility step. These results nicely agree with numerically determined exact solutions even beyond the limit of small activities.  We further evaluated the effect of non-zero curvature of the active-passive interface on the polarization and density. Reduced polarization is smaller than for planar activity steps, whereas the contrast in the bulk densities is smaller/larger for concave/convex active-passive interfaces. Both the maximum polarization and the bulk density ratio depend on the curvature of the interface. 

Furthermore, we derived an exact formula for the (radial) total polarization induced by an arbitrary radially symmetric activity landscape. Compared to the result for onedimensional activity landscapes, the total polarization contains a non-local, geometry-induced correction. The total polarization is thus no longer determined solely by bulk variables. This result proves that curved active/passive interfaces generally yield a geometry-induced contribution to the emergent density and polarization profiles, and the associated total polarization and bulk density ratio.

  \section*{Acknowledgements}
\label{sec:acknowledgements}

We thank Klaus Kroy for valuable discussions.
We acknowledge funding by the Deutsche Forschungsgemeinschaft (DFG) through the priority program “Microswimmers” (SPP 1726, project 237143019), and by Czech Science Foundation (project No. 20-02955J).
Viktor Holubec gratefully acknowledges support by the Humboldt foundation.

\bibliography{references}

\appendix

\section{Polarization peak and density ratio}
\label{sec:circular-setup}

Within the active ($r < \rif$) and passive region ($r > \rif$) the general solution of the respective polarization and density profiles read [Eqs.~\eqref{eq:pol_circ_nat_bound} and \eqref{eq:rho_act_circ_nat_bound}] 
\begin{align}
  \label{eq:polarization_active_circ}
  \Pa(r)
  &=
    \Ca I_1(r/\la),
    \quad
    \la
    =
    \left(
    \frac{\Dr}{D}
    +
    \frac{v_0^2}{2D^2}
    \right)^{-1/2},
  \\[0.5em]
  \label{eq:polarization_passive_circ}
  \Pp(r)
  &=
    \Cp K_1(r/\lp),
    \quad
    \lp
    =
    \sqrt{\frac{D}{\Dr}},
  \\[0.5em]
  \label{eq:density_active_circ}
  \rho_{\te a}(r)
  &=
    \rho_{\te a}
    +
    \frac{\Ca \la v_0}{D}
    \left[
    I_0(r/\la)
    -
    1
    \right],
  \\[0.5em]
  \label{eq:density_passive_circ}
  \rho_{\te p}(r)
  &\equiv
    \rho_{\te a}(\rif)
    \equiv
    \rho_{\te p}
    = const.,
\end{align}
with $\rho_{\te a} \equiv \rho_{\te a}(0)$. The emerging integration constants $\Ca$ and $\Cp$ follow from the continuity conditions
\begin{equation}
  \label{eq:matching_continuity_circ}
  \Pp(\rif)
  =
  \Pa(\rif),
  \quad
  \Pa'(\rif) - \Pp'(\rif)
  =
  \frac{v_0}{2D} \rho(\rif),
\end{equation}
given Eqs.~\eqref{eq:matching_intuit} and \eqref{eq:jump_condition} in the main text.
The first condition implies that
\begin{equation}
  \label{eq:ca_in_terms_of_cp_circ}
  \Cp
  =
  \Ca
  \frac{I_1(\rif/\la)}{K_1(\rif/\lp)}.
\end{equation}
For the sake of brevity, we introduce the abbreviations
\(
I_n \equiv I_n(\rif/\la)
\)
and
\(
K_n \equiv K_n(\rif/\lp).
\)
The second condition in Eq.~\eqref{eq:matching_continuity_circ} yields
\begin{align}
  \Pa' - \Pp'
  =
  \Ca \frac{I_0 + I_2}{2\la}
  +
  \Cp \frac{K_0 + K_2}{2\lp}
  =
  \frac{v_0}{2D} \Rp
\end{align}
and using Eq.~\eqref{eq:ca_in_terms_of_cp_circ} we get
\begin{equation}
  \label{eq:Ca}
  \frac{\Ca}{\Rp}
  =
  \frac{v_0}{2D}
  \left(
    \frac{I_0 + I_2}{2\la}
    +
    \frac{I_1}{K_1}
    \frac{K_0 + K_2}{2\lp}    
  \right)^{-1}.  
\end{equation}
Knowing that the polarization peaks exactly at the interface (see Fig.~\ref{fig:pol_rho_th_vs_num}), it follows from Eqs.~\eqref{eq:polarization_active_circ} and \eqref{eq:density_passive_circ} that the maximum relative polarization $p(\rif)/\rho(\rif)$ is given by $\Ca I_1/\Rp$.
Using Eq.~\eqref{eq:Ca} and introducing the geometry function~\eqref{eq:geometry_fct},
\begin{equation}
  \label{eq:geometry_fct_app}
  G_\te{A}(\rif)
  =
  \left(  
    \frac{I_0+I_2}{2\la I_1}
    +
    \frac{K_0+K_2}{2\lp K_1}    
  \right)^{-1},
\end{equation}
the maximum relative polarization reads
\begin{align}
  \label{eq:Pmax_circ_app}
  \frac{p(\rif)}{\rho(\rif)}
  &=
    \frac{v_0}{2D}
    G_\te{A}(\rif),
\end{align}
as given in Eq.~\eqref{eq:Pmax_circ}.
The relative density profile,
\(
(\rho - \Ra)/\Rp
\), and thus also the bulk density ratio, $\Rp/\Ra$, can be obtained using Eqs.~\eqref{eq:density_active_circ}, \eqref{eq:density_passive_circ}, and \eqref{eq:Ca} and a similar approach.

\section{Interior passive -- Exterior active}
\label{sec:reversed-situation}

We now consider the case where the particle is passive for $r < \rif$, and otherwise active. In analogy to the calculations in App.~\ref{sec:circular-setup} one now has

\begin{align}
  \label{eq:polarization_active_circ}
  \Pp(r)
  &=
    \Cp I_1(r/\lp),
  \\[0.5em]
  \label{eq:polarization_passive_circ}
  \Pa(r)
  &=
    \Ca K_1(r/\la),
  \\[0.5em]
  \label{eq:density_passive_circ}
  \rho_{\te p}(r)
  &\equiv
    \rho_{\te a}(\rif)
    \equiv
    \rho_{\te p}
    = const.,
  \\[0.5em]
  \label{eq:density_active_circ}
  \rho_{\te a}(r)
  &=
    \rho_{\te p}
    +
    \frac{\Ca \la v_0}{D}
    \left[
    K_0(\rif/\la)
    -
    K_0(r/\la)
    \right].
\end{align}
Note that the argument of $I_1$ ($K_1$) now carries $\lp$ ($\la$) as characteristic length scale. Integration constants $\Ca$ and $\Cp$ are determined by the same continuity conditions \eqref{eq:matching_continuity_circ} as in App.~\ref{sec:circular-setup}, yielding
\begin{align}
  \label{eq:ca_in_terms_of_cp_2}
  \Cp
  &=
    \Ca
    \frac{K_1(\rif/\la)}{I_1(\rif/\lp)},
  \\[0.5em]
  \label{eq:Ca_2}  
  \frac{\Ca}{\Rp}
  &=
  -\frac{v_0}{2D}
    \left(
    \frac{K_1}{I_1}
    \frac{I_0 + I_2}{2\lp}
    +
    \frac{K_0 + K_2}{2\la}    
  \right)^{-1},
\end{align}
where we used the abbreviations
\(
I_n \equiv I_n(\rif/\lp)
\)
and
\(
K_n \equiv K_n(\rif/\la).
\)
Reduced density and polarization profiles, $\rho/\Rp$ and $p/\Rp$, are plotted in Fig.~\ref{fig:pol_rho_th_vs_num} c) and d).
Introducing the geometry function
\begin{equation}
  \label{eq:geometry_fct_p}
  G_\te{P}(\rif)
  \equiv
  \left(  
    \frac{I_0+I_2}{2\lp I_1}
    +
    \frac{K_0+K_2}{2\la K_1}    
  \right)^{-1},
\end{equation}
the (negative) polarization peak and the bulk density ratio are given by
\begin{align}
  \label{eq:Pmax_circ_p}
  \frac{p(\rif)}{\rho(\rif)}
  &=
    -\frac{v_0}{2D}
    G_\te{P}(\rif),
  \\[0.5em]
  \label{eq:density_ratio_circ_p}
  \frac{\Ra}{\Rp}
  &=
    1
    -
    \frac{v_0^2 \la}{2 D^2}
    G_\te{P}(\rif)
    \frac{K_0}{K_1}.
\end{align}
Both quantities are plotted in Fig.~\ref{fig:pol_rho_th_vs_num} e) and f) (dotted lines). For $\rif \to \infty$, both approach their counterparts at straight planar active-passive interfaces [Eqs.~\eqref{eq:Pmax_limit} and \eqref{eq:density_ratio_limit}].

\end{document}